\documentclass[reprint,cha,superscriptaddress]{revtex4-1}

\usepackage{graphicx}
\usepackage{amssymb}
\usepackage{placeins}
\usepackage{fancyhdr}
\usepackage{nameref,hyperref}

\usepackage[T1]{fontenc}
\usepackage{color}
\usepackage{url}
\usepackage{rotating}
\usepackage[squaren]{SIunits}
\usepackage{amsmath}
\usepackage{wasysym}
\usepackage[latin1]{inputenc}
\usepackage[T1]{fontenc}
\usepackage{ae,aecompl}
\usepackage{xcolor}

\usepackage[font=footnotesize,format=plain,justification=justified,singlelinecheck=false]{caption}
\newcommand{\eqna}{\begin{eqnarray}}
\newcommand{\eqne}{\end{eqnarray}}
\newcommand{\eqnaa}{\begin{align*}}
\newcommand{\eqnee}{\end{align*}}

\definecolor{darkgreen}{rgb}{0,0.5,0}

\begin{document}

\title{The Contribution of Different Electric Vehicle Control Strategies to Dynamical Grid Stability}

\author{Sabine Auer}
 \email{auer@pik-potsdam.de}
 \affiliation{Potsdam Institute for Climate Impact Research, 14412 Potsdam, Germany}
 \affiliation{Department of Physics, Humboldt University Berlin, 12489 Berlin, Germany}
\author{Casper Roos}%
\affiliation{Potsdam Institute for Climate Impact Research, 14412 Potsdam, Germany}\author{Jobst Heitzig}
\affiliation{Potsdam Institute for Climate Impact Research, 14412 Potsdam, Germany}
\author{Frank Hellmann}%
\affiliation{Potsdam Institute for Climate Impact Research, 14412 Potsdam, Germany}%
\author{J\"urgen Kurths}
\affiliation{Potsdam Institute for Climate Impact Research, 14412 Potsdam, Germany}
\affiliation{Department of Physics, Humboldt University Berlin, 12489 Berlin, Germany}
\affiliation{Institute of Applied Physics, Russian Academy of Science, 603950 Nizhny Novgorod, Russia}

\markboth{%Journal of \LaTeX\ Class Files,~Vol.~14, No.~8, August~2015
}%
{Auer \MakeLowercase{\textit{et al.}}: Electric Vehicle Control Strategies for Dynamical Grid Stability}
% The only time the second header will appear is for the odd numbered pages
% after the title page when using the twoside option.
% 
% *** Note that you probably will NOT want to include the author's ***
% *** name in the headers of peer review papers.                   ***
% You can use \ifCLASSOPTIONpeerreview for conditional compilation here if
% you desire.

% If you want to put a publisher's ID mark on the page you can do it like
% this:
%\IEEEpubid{0000--0000/00\$00.00~\copyright~2015 IEEE}
% Remember, if you use this you must call \IEEEpubidadjcol in the second
% column for its text to clear the IEEEpubid mark.

% use for special paper notices
%\IEEEspecialpapernotice{(Invited Paper)}

% make the title area
%\maketitle

% As a general rule, do not put math, special symbols or citations
% in the abstract or keywords.
\begin{abstract}

A major challenge for power grids with a high share of renewable energy systems (RES), such as island grids, is to provide frequency stability in the face of renewable fluctuations. In this work we evaluate the ability of electric vehicles (EV) to provide distributed primary control and to eliminate frequency peaks. To do so we for the first time explicitly model the network structure and incorporate non-Gaussian, strongly intermittent fluctuations typical for RES.

We show that EVs can completely eliminate frequency peaks. Using
threshold randomization we further demonstrate that demand synchronization effects and battery stresses can be greatly reduced. In contrast, explicit frequency averaging has a strong destabilizing effect, suggesting that the role of delays in distributed control schemes requires further studies.

Overall we find that distributed control outperforms central one. The results are robust against a further increase in renewable power production and fluctuations.

%Power grids with high shares of fluctuating renewable energy sources (RES) such as island grids limit their RES penetration for reasons of dynamic grid stability. With this work on primary EV control we contribute to the variety of novel power grid control techniques to overcome such issues.
%
%The combination of explicitly modeling the network structure and incorporating the true intermittent nature of fluctuations from RES is novel for such a grid control study.
%
%In our study, we identified a combination of minimal necessary ramping slope and randomized battery threshold as, to our knowledge, best to jointly ensure grid stability and avoid battery degradation. In comparison, an input signal averaging led to a complete grid frequency destabilization instead of avoiding the synchronization of appliances. With regard to its effectiveness, decentral outperformed central control.
%This model setup is robust against a further increase in power production and thus fluctuations. 
\end{abstract}

% Note that keywords are not normally used for peerreview papers.
%\begin{IEEEkeywords}
%IEEE, IEEEtran, journal, \LaTeX, paper, template.
%\end{IEEEkeywords}

\maketitle

% For peer review papers, you can put extra information on the cover
% page as needed:
% \ifCLASSOPTIONpeerreview
% \begin{center} \bfseries EDICS Category: 3-BBND \end{center}
% \fi
%
% For peerreview papers, this IEEEtran command inserts a page break and
% creates the second title. It will be ignored for other modes.
%\IEEEpeerreviewmaketitle

With the increasing share of intermittent renewable energy sources (RES) in Germany, it is becoming more challenging to maintain the dynamical stability of power grids~\cite{schafer2016taming,troester2009new}. Intermittent RES have strong power output fluctuations on a short time scale which cause imbalances between the power production and consumption~\cite{rohden2015synchronization,short2007stabilization}. To maintain the grid's dynamical stability, the authors of \cite{schafer2015decentral} suggested the concept of Decentral Smart Grid Control (DSGC), where power consumers adjust their demand according to the locally measured grid frequency. The use of the locally measured grid frequency for DSGC has the advantage that the electrical appliances can be automated with load controllers, such as the Distributed Intelligent Load Controller (DILC)~\cite{short2007stabilization,ian2003intelligent}. These load controllers then adjust the power demand of an electrical device with a certain control strategy or heuristic~\cite{short2007stabilization}. 

Today, in Germany around 90\% of RES are installed in distribution grids \cite{verteilnetzstudie}, the lower grid levels of the hierarchical power grid infrastructure. In order to balance fluctuations locally where they appear, electric vehicles (EV) and their battery storage systems would present an ideal use case for DSGC~\cite{liu2013decentralized,pillai2010vehicle}. EVs can adjust their power demand within milliseconds and have the capability to deliver power back into the grid, also known as \emph{vehicle to grid} (V2G) power transfer and vice versa simply charge -- \emph{grid to vehicle} ~\cite{pasaoglu2013projections,wang2011impact}. 

With a frequency control strategy, EVs essentially act as primary frequency control reserves, since they autonomously assist in stabilizing the grid frequency~\cite{pillai2010vehicle,liu2013decentralized}. The fact that 94\% of all U.S cars are parked at noon time of a typical day \cite{van2011assessment} shows the great potential for EV control. Instead of installing additional expensive balancing hardware, the anyways idle EVs may be used for grid control purposes.

A control strategy that has been suggested and used in order to maintain dynamical grid stability is the \emph{band gap} strategy~\cite{liu2013decentralized,almeida2011electric,karki2014reliability}. 
In this control strategy a dead-band, the frequency interval between the battery thresholds for positive and negative frequency deviations, is predefined where no power, relative to the base charging scenario, is transferred between the EV and the power grid, as small frequency deviations are considered to be part of normal operation. When the frequency deviations are out of this band gap, the EV and power grid exchange an amount of power that depends on the magnitude of the deviation from the band gap and a predefined rate of power transfer called the ramping rate. Thus, this rate of power transfer and the frequency band gap are the parameters which determine the sensitivity of this control strategy. The performance of this control strategy is evaluated with regard to its grid stability improvement and the number of battery switchings. \emph{Switching events} include the battery action changes: decharging to idle,  charging to idle, idle to charging and idle to decharging. The grid stability is evaluated with regard to the \emph{threshold exceedance} which is the time share the frequency spends outside a given safe band.

In this study, we aim to find an optimal parameterization for the EV control in a modeling scenario with very strong Photovoltaic (PV) penetration \cite{Roos2016}. Our parameterization is such that it improves the grid's dynamical stability, minimizes the amount of switching events to avoid battery degradation, and ensures an effective control at the same time. 
In addition, we test whether randomization will be useful in order to prevent undesirable demand synchronization as observed in \cite{krause2015econophysics,short2007stabilization,mohsenian2010autonomous,moghadam2016distributed}.

In contrast to previous works, we explicitly model the network structure as a complex network which gives us the opportunity to investigate the influence of decentral vs. central control and model the interaction of appliances via the power grid. In addition, stochastic models reproducing solar power fluctuations are very recent \cite{anvari2016short} and we are the first to incorporate the true intermittent nature of fluctuations from RES in such a grid control study.

As a simplification, all nodes in our network have the same absolute power and inertia to exclude any side effects from network heterogeneities that make the evaluation of different control strategies more difficult.

\par
For our model setup, by numerical simulations we find a minimal necessary (critical) ramping rate that completely suppresses threshold exceedance and therefore improves grid stability. We reproduced the relation between ramping rate and frequency deviation and thus exceedance analytically. The same ramp for all EVs leads, as expected, to a synchronization of the control devices with the result of
 a large number of battery switching events. Thus, we complete the EV control scheme with a randomization approach and allow for a variance in battery threshold. We identify this combination of ramping slope and randomized battery threshold as, to our knowledge, best to jointly reduce exceedance and switching events. In comparison, an approach, that uses time averaged frequency input signals for the EV control, did not show the same success in switching event reduction but even destabilized the power grid.
Our identified best control parameterization works over a wide  range of power production and thus for different fluctuations strength. We find that switching events only increase slightly. 
Regarding the question of central vs. decentral control, we identify decentral control to be more effective.

This paper is organized as follows. First, we introduce our model and the chosen grid parameterization, the modeling of the inverter dynamics and the EV integration, and we present the stability methods to evaluate different parameterization scenarios. In the Results section we start by investigating a base scenario without EV control. Then we test different EV parameterizations or control strategies and their robustness. We close the results section with a comparison of decentral vs central control. Finally, we draw conclusions and discuss open questions and future work.

\section*{Model and Stability Methods}

\begin{figure}[h!]
\centering
\includegraphics[width=0.3\textwidth]{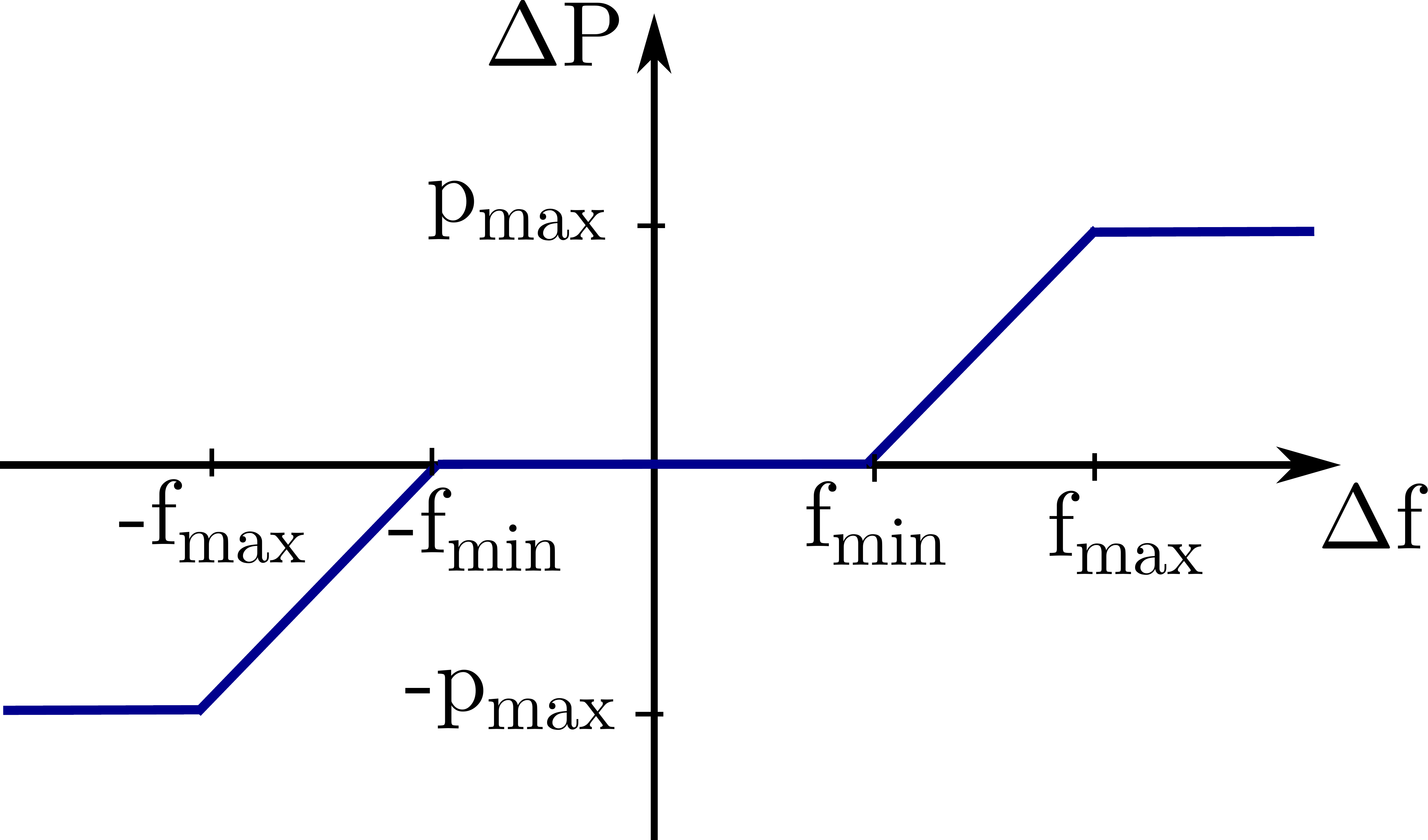}
\caption{\textbf{Scheme of battery ramping}. $[-f_{min},f_{min}]$ is the frequency dead band for which the battery stays idle, $\Delta P=0$, no frequency control is provided. $f_{min}$ and $f_{max}$ determine the ramp at which the battery charges ($\Delta P >0$) or discharges ($\Delta P >0$). $\pm p_{max}$ are the upper limits for charging and discharching.}
\label{fig:ramping_scheme}
\end{figure}

Our model is designed to represent distribution grids. Thus, we chose tree-shaped networks as the underlying network topology (generated with a random growth model \cite{schultz2014random}) and introduce lossy lines, since the common assumption of non-lossy lines for transmission grids does not hold for distribution grids. The model uses time steps of 0.01 seconds, and each control strategy was simulated on 15 different power grids, using a Monte Carlo simulation in order to average out the influence of the network structure.

In the following, we elaborate on the modeling assumptions concerning the type of node dynamics as well as the Mid-Voltage (MV) grid and EV parameterization. Then, we describe what measures we use to evaluate the performance of different heuristics.

\paragraph{Inverter Dynamics} 
Our choice of a MV grid region with high PV penetration make the grid dynamics inverter-dominated. Most PV and wind power plants are connected to the grid via grid-feeding inverters, however, grid-forming inverters are important for dynamic stability \cite{schiffer2016survey}. Thus in our scenario, which is meant to represent future MV grids, we assume that effective grid nodes representing an accumulation of production from the lower Low-Voltage (LV) levels where each node has at least one grid-forming inverter. This type of inverter is able to provide virtual inertia whereas grid-feeding inverters contribute no inertia. The classical power grid model (or swing equation) is derived from the Synchronous Machine Model representing conventional generators and their rotating masses \cite{nishikawa2015comparative}. Grid-forming inverters and their power electronics may be programmed as Virtual Synchronous Machines, as mentioned before, by using a smooth droop control. This then leads to the same equations for the voltage angle $\phi$ and frequency $\omega=2\pi f$ in terms of the (virtual) inertia $H$ \cite{schiffer2013synchronization}, power infeed $P$, (virtual) damping $\alpha$, line susceptibilities, $Y=G+jB$, and voltage magnitudes $U$ for each node $i$:
\begin{align}
\begin{split}
\dot{\phi}_i=&\omega_i,\\
\dot{\omega_i}=&\frac{1}{H_i}(P_i+\delta P_i(t)-\alpha\omega_i
-b(f_i)\\
 &-\sum_kU_i|Y_{ik}|U_k \sin(\phi_i-\phi_k+\phi_{ik})).
\end{split}
\label{eq:dynmics}
\end{align}
where $b(f)$ is the function of the bandgap strategy illustrated in Fig. \ref{fig:ramping_scheme} and equals
\begin{equation}
b(f_i)=\Theta(|f_i|-f_{min})sign(f_i)(|f_i|-f_{min})r
\end{equation}
with the Heaviside step function $\Theta$ and the sign function $sign$.
The power ramping slope of the EV batteries is given by 
\begin{equation}
r = \frac{p_{max}}{f_ {max}-f_{min}}
\end{equation}
where $p_{max}= 3.7$kW is the maximum charging power.

The virtual inertia and damping for the network model is given by the low-pass filter exponent $\tau_p$ and the droop control parameter $k_p$ from grid-forming inverters: $H_i = \tau_p/k_p$, $\alpha_i=1/k_p$, $\forall i$ with $i=1,..,N$. 
Standard parameters for the droop and time constants of grid-forming inverters are in the range $\kappa_p=[0.1,..,10]$ and $\tau_p=[0.1,...,10]$ \cite{schiffer2013synchronization,coelho2002small}.
As we are interested in the low inertia case, with few low powered grid forming inverters at each node, we assume a weakly reacting, strongly smoothed system. This leads us to consider $\alpha=0.01s$ and $H=0.05s^2$. We note that the results are not sensitive to the exact choice of $\alpha$ and $H$.

%\begin{figure}[t]%
%\begin{minipage}[c]{0.5\columnwidth}
%\hspace{-0.2\textwidth}
%    \includegraphics[width=0.8\columnwidth]{freq_stoch_batch0_disp3_net12_edited_bw.pdf}
%      \vspace{-0.1\textwidth}
%  \end{minipage}\hspace{-0.04\textwidth}
%  \begin{minipage}[c]{0.55\columnwidth}
%    \caption{\textbf{Base Scenario time series:} Time series of all nodes' frequency for the base scenario of an example grid with power production of $0.268$ MW and $0.168$ MW demand. The exceedance is calculated as the time share of frequency trajectories outside the grey safety band of $\pm 20$ mHz. The dark grey dotted line at $\pm 10$ mHz marks the threshold for battery control.}
%     \label{fig:time_series_basescenario}
%  \end{minipage}
%  \vspace{0.0\textwidth}
%\end{figure}%

\begin{figure}[t!]
\centering
\includegraphics[width=0.28\textwidth]{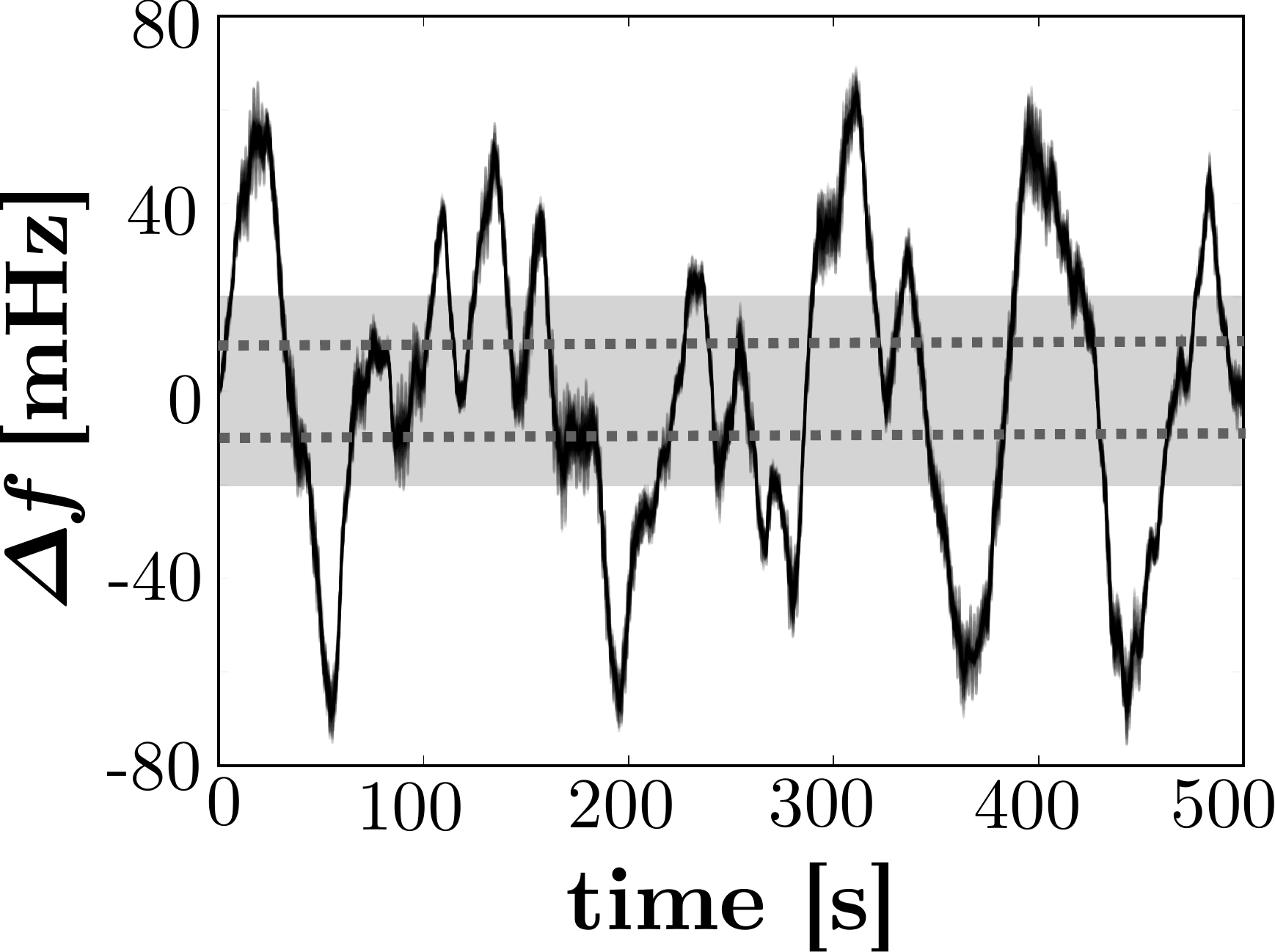}
\vspace{0.01\textwidth}
\caption{\textbf{Base Scenario time series:} Time series of all nodes' frequency for the base scenario of an example grid with power production of $0.268$ MW and $0.168$ MW demand. The exceedance is calculated as the time share of frequency trajectories outside the grey safety band of $\pm 20$ mHz. The dark grey dotted line at $\pm 10$ mHz marks the threshold for battery control.}
\label{fig:time_series_basescenario}
\end{figure}

\paragraph{Mid-Voltage (MV) grid Parameterization} The MV grid is a good testing case for modeling EV frequency control as a reaction to power fluctuations caused by a high PV penetration. This is the case because most PV power plants are connected to low-voltage (LV) or MV levels. In this modeling scenario, we chose a network of 100 nodes. It thereby represents an average German MV grid because Germany has 4,500 MV distribution networks that connect 500,000 LV distribution networks \cite{bossmann2015shape}. 
All nodes have the same amount of inflexible load and production which a strong assumption in favor of homogeneity that allows to attribute any difference in performance of EV control at different network nodes purely to the chosen control strategy in combination with the nodes' network properties. 

For the inflexible load and average PV power generation a challenging 2050 scenario was assumed, where the power production from PV is two times larger than the inflexible load in the MV nodes. Here, we assume $0.268$MW solar production for each MV node. This is a challenging, but realistic scenario, as the installed PV capacity in some LV grids in south Germany can already exceed the peak load by a factor of ten \cite{von2013time}. The inflexible load of each node was $0.168$MW, as the peak load in 2014 of $84$ GW in the German grid was equally divided among the MV nodes \cite{bayer2015report}. This peak load is assumed to remain unchanged until 2050, although it included the
additional load from EVs. This is the case since Smart Charging of EVs and the improved energy efficiency are expected to compensate for additional loads from the growing amount of EVs \cite{bossmann2015shape}. Hence, the effective power input $P_i$, the power which is injected into the grid, equals $P_i=0.1$MW. 

For simplicity and homogeneity, all $99$ MV nodes then also have the same inertia. 
For the representation of the upper grid levels there is one heavy node (slack bus, labeled as node $0$) responsible for power balance with a power input built from the negative sum over all MV nodes' power in-feeds and losses on the lines:
$P_0 = \sum_{i=1}^N P_i +P_{loss}$. As the name ``heavy node'' tells, the slack bus' inertia highly exceeds the lower level nodes' inertia, here we assume: $H_0=\sum_{i=1}^{100} H_i$.

The impedance of the lines for typical Mid-Voltage grid lines with $20$kV base voltage equals $Z= Y^{-1}=(G+iB)^{-1} = (0.4 + 0.3j) \Omega/km$ \cite{auer2016can}. The coupling strength between a node pair $(i,j)$ then equals $K_{ij}=U_i|Y_{ij}|U_j$. The addition of the resistance leads to line losses and at the same time introduces a phase shift of $\phi_{ij}\approx \arctan(\frac{G_{ij}}{B_{ij}})$ which was shown to have significant consequences for stability \cite{auer2017stability}.

\paragraph{EV parameterization}
The EV's maximum charging/injection power transfer rate is assumed to be 3.7 kW (230V/16A), also referred to as private home charging, since this type of EV charging is expected to have a market share of 64,8\% in Germany by 2050 \cite{madina2016methodology,richter2010potenziale}.
%In the system model, no delay time was included, as current literature on DSGC has so far not included delay times into their models. Thus, to make the results of this study comparable with previous studies it was assumed that the load controller had infinitely fast measuring and calculation equipment. 
The total battery capacity of an EV was 90 kWh, equal to the maximum capacity of a Tesla model S \cite{tesla}. The energy consumption during a driving event was 6.7 kW, by assuming the average speed of the New European Driving Cycle of 33.6 km/h and the average power consumption of 0.2 kWh/km for small and medium sized EVs \cite{metz2012electric,silva2009analysis}. At the beginning of each simulation 94\% of the EVs were available, in compliance with the findings from \cite{van2011assessment},which documented that 94\% of all U.S cars were parked at noon on a typical day. The 6\% of unavailable EVs were randomly distributed among the MV nodes in the model. %and assumed to be driving with a driving time drawn a gamma probability distribution function, as observed by several empirical studies \cite{cao2013methodology,hutchinson2012car}. The gamma PDF is given by
%\begin{equation}
%p(x) = x^{k-1}\frac{e^{-x/\Theta}}{\Theta^k \Gamma(k)}
%\label{eq:gamma_pdf}
%\end{equation}
%where $x$ is the driving duration, $k$ a positive integer shape parameter, $\Theta$ the scale parameter and $\Gamma(k)$ the gamma function.
 
The initial charge of all EVs is 72 kWh representing a state of charge (SOC) of 80\%. For the EV battery threshold, we assume: $f_{min}=0.01$ (see Figs. \ref{fig:ramping_scheme}\&\ref{fig:time_series_basescenario}), which corresponds to the so-called dead band from the German transmission code and defines at which frequency primary control actions kick in to balance deviations from the desired $50$Hz set point \cite{transmission_code}.

\begin{figure}[b!]
\centering
\includegraphics[width=0.34\textwidth]{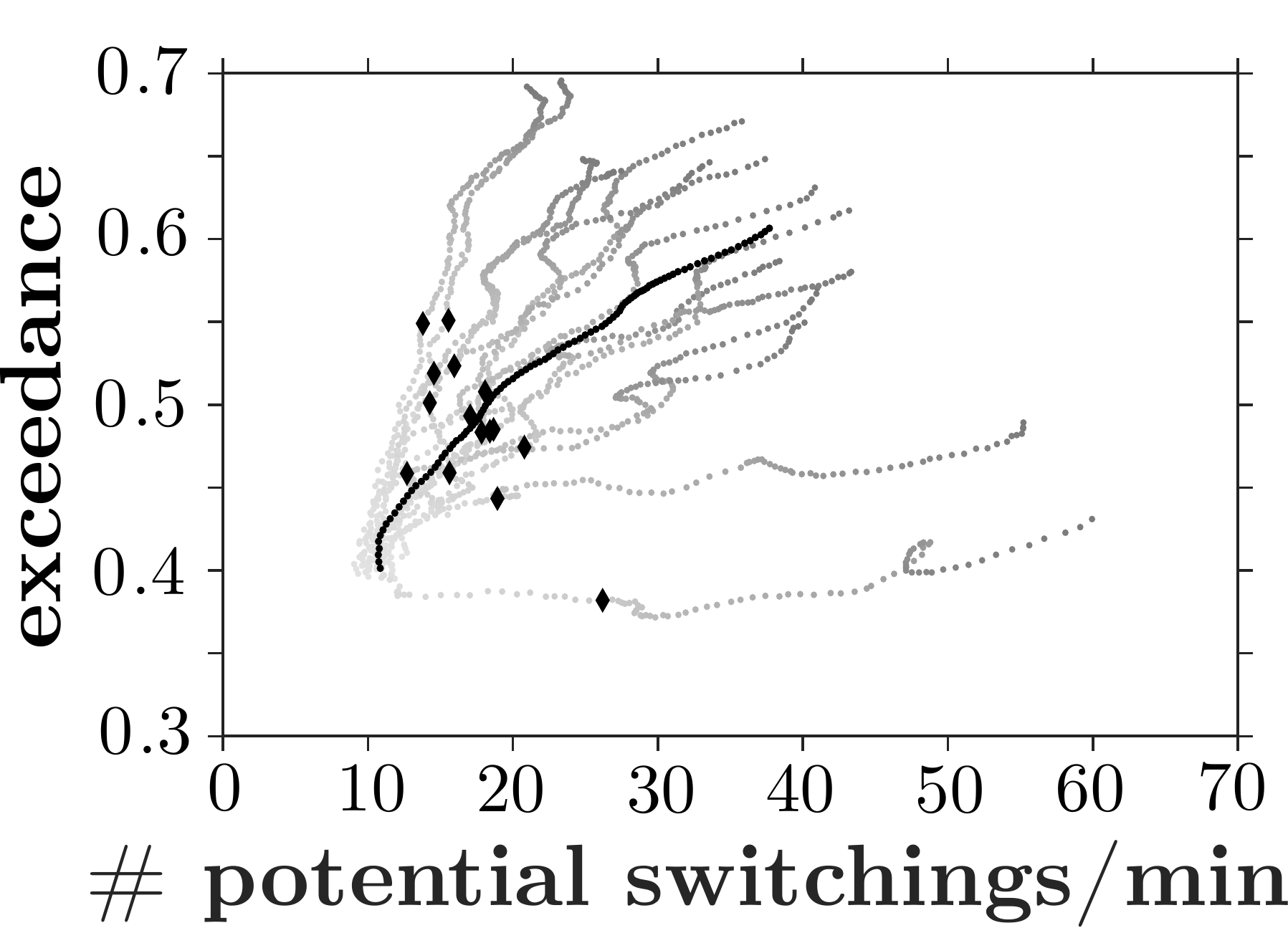}
%\hspace{0.01\textwidth}\includegraphics[height=0.16\textwidth]{freq_stoch_batch0_disp3_net12_edited_bw.pdf}
\vspace{0.01\textwidth}
\caption{\textbf{Sensitivity of Base Scenario (no EV Control):} Exceedance (averaged over all nodes) plotted over the potential average number of battery switchings for increasing power production from $0.17$ to $0.5$ MW. The grey dotted lines (darker grey for greater power production) show a network sample of 15 random Mid-Voltage (MV) topologies with the black line representing the ensemble average. The black diamonds mark the chosen base scenario. }
\label{fig:power_increase_no_battery}
\end{figure}

\begin{figure*}[t!]
\centering
\includegraphics[height=0.2\textwidth]{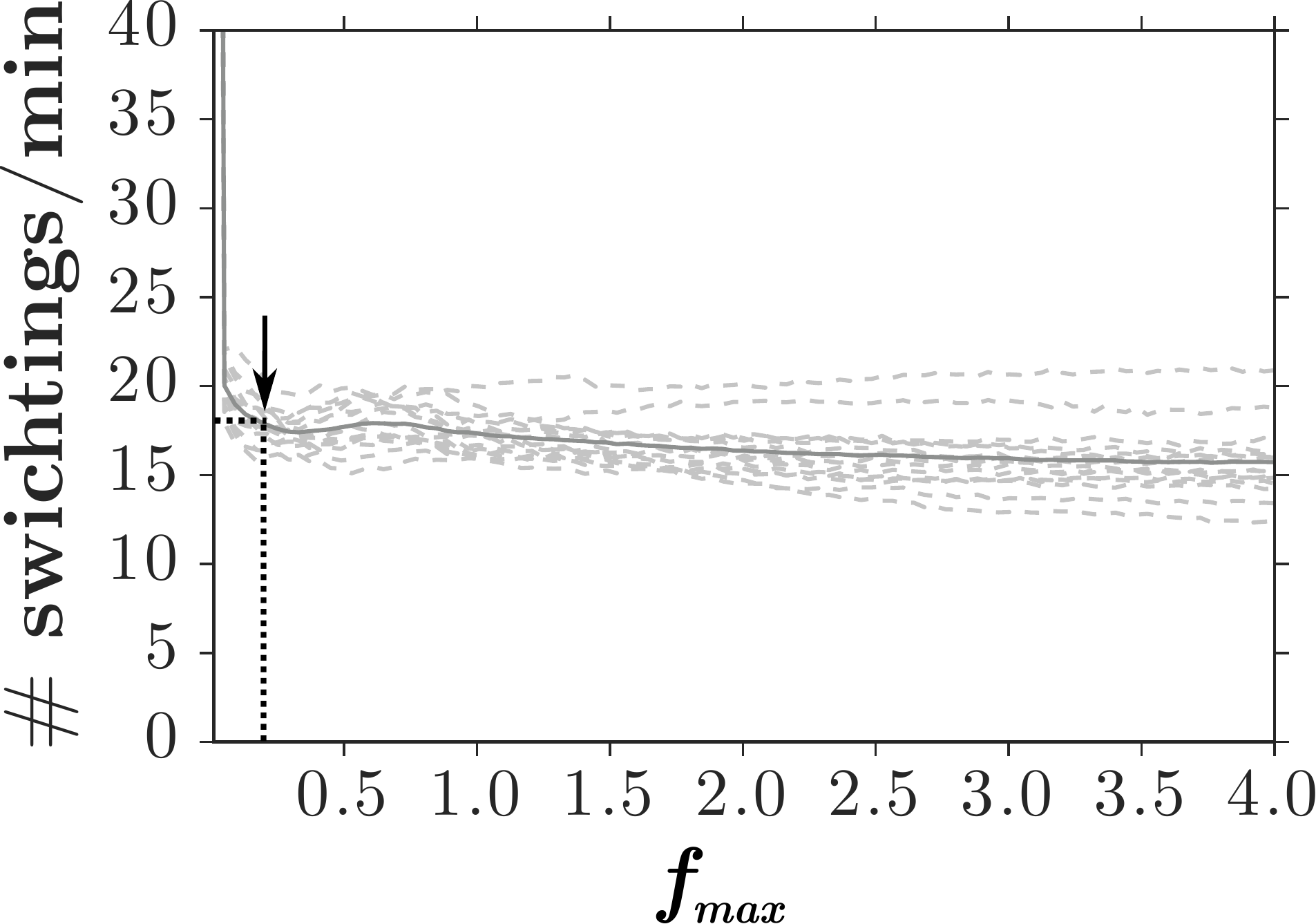}
\hspace{0.02\textwidth}
\includegraphics[height=0.2\textwidth]{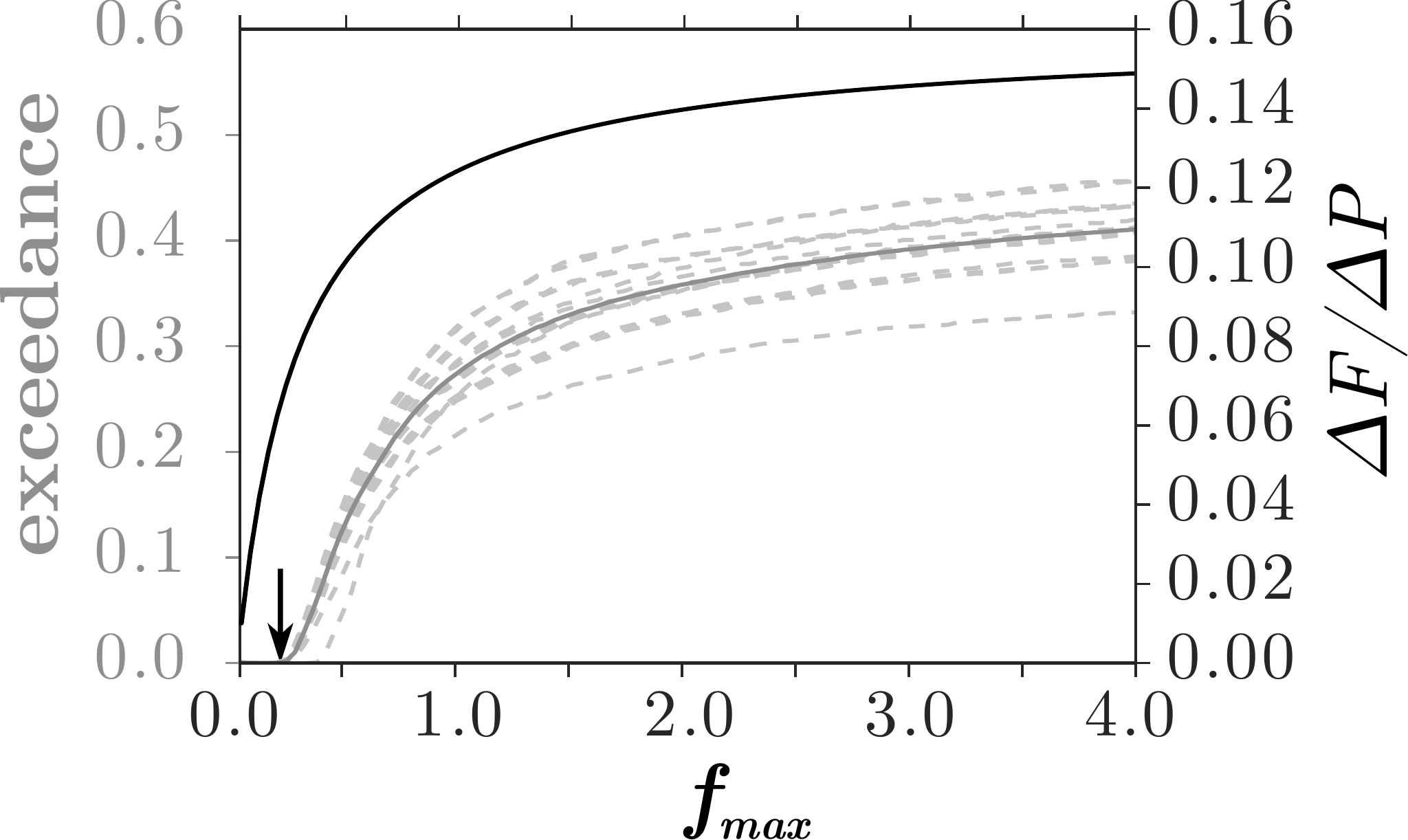}
\hspace{0.02\textwidth}
\includegraphics[height=0.2\textwidth]{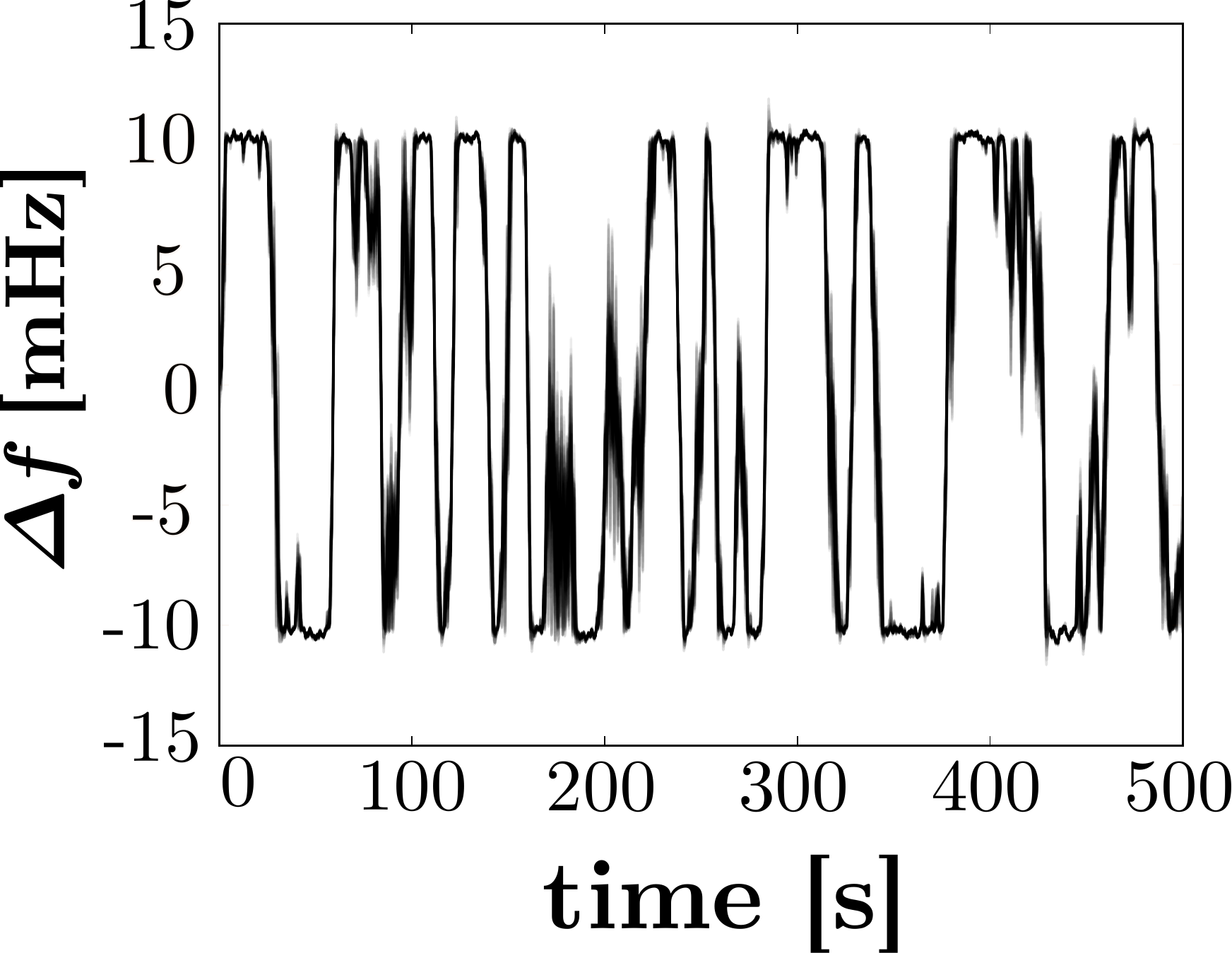}
\caption{\textbf{Parameterization of EV ramping}: Average number of nodal battery switchings (left) and average nodal exceedance (center) plotted over different ramping slopes $r$ for $15$ different networks (grey dashed lines). For $f_{max}=0.01$ switching events go up to $500$. The solid grey line represents the ensemble average. The second y-axis of the center plot shows the analytic result for global frequency offset $\Delta F$ caused by a power change $\Delta P$ (see \eqref{eq:DeltaF}). Right: Frequency time series for $f_{max}=0.02$.}
\label{fig:ramping_slope_increase}
\end{figure*}

\paragraph{Stability Measures}
The stability measures typically used in power grid synchronization analysis are not applicable to our stochastic system \cite{hellmann2015survivability,menck2013basin,
nishikawa2006synchronization,pecora1998master,belykh2004connection,Auer2016}. 
Linear stability of a particular operational state, assumed to be a fix point of the grid model dynamic equations, against small perturbations is given by the largest non-zero eigenvalue of the linearized dynamics around the fix point \cite{nishikawa2006synchronization,pecora1998master,belykh2004connection}. However, problems arise for larger disturbances or if the Laplacian is not diagonalizable \cite{Wei2011}, e.g. if the ohmic resistances of transmission lines are not neglected.
There are methods for the assessment of a dynamic system even against large perturbations that rely on a sampling-based approach.
The global stability of a fixed point of a dynamical system can be quantified by the volume of its basin of attraction. Then, a system's basin stability equals the probability to asymptotically return to the stable point of operation after an initial perturbation \cite{menck2013basin}. 
Survivability measures the ability of a system to keep within some predefined operating regime when experiencing large perturbations \cite{hellmann2015survivability}.
The previous stability measures are mostly studied for deterministic systems. Though first generalizations of Basin stability to stochastic systems have appeared \cite{zheng2016transitions,serdukova2016stochastic}.

Here, we use the {\it exceedance} to quantify the stability of the synchronous state. It is the fraction of time an observable stays outside a defined ``safe'' region. For our case we define a frequency threshold of $0.02$ Hz (see Fig. \ref{fig:time_series_basescenario}).
We apply fluctuations to all network nodes but the slack bus and then record the frequency response for each node $i=1,..,N$. Thus, we end up with $N$ frequency time series from which stability measures are derived, namely the exceedance $E_i$ or probability $p_i$ for each node $i$ to exceed $20$ mHz.
\begin{equation}
E_{i}=p_i(|f_i|>0.02~\text{Hz}).
\label{eq:exc}
\end{equation}
This can be further aggregated into the average exceedance over all $N$ nodes: 
% Drivers of Mean Exceedance, dominance, troublemaker index, infectiousness
\begin{equation}
\bar{E}=\frac{1}{N}\sum_{i=1}^N E_{i}
\label{eq:ti}
\end{equation}

Besides frequency stabilization the performance of the proposed heuristics are evaluated with respect to their influence on battery degradation. Hence, the battery switching events are recorded. Switching events include the battery action changes: decharging to idle,  charging to idle, idle to charging and idle to decharging. Note that we evaluate switching only for the primary control. However, the background charging for battery refilling is assumed to be fulfilled in our power balance and considered to be a problem of secondary or tertiary control.

% Results and Discussion can be combined.
\section*{Results}

The starting point of our investigations is the base scenario in order to gather an understanding of how our power system behaves with increasing power production from RES without any EV primary control present. In the Model Section, we reasoned to choose a modeling scenario with an amount of intermittent RES production that provides a challenging base scenario for our EV DSGC.  Nevertheless, at first we want to better understand how larger or lower values of RES production influence the power system stability measures. Then, we identify the battery ramping slope necessary to prevent frequency from exceeding the chosen safety margin of $\pm0.02$Hz. In order to identify not only a grid- but also battery-friendly control mechanism, we apply a suitable battery threshold randomization. We compare this strategy with the alternative approach of averaging over past frequency values in order to overcome fast switching. Finally, the importance of decentral control is investigated more closely with respective to their effectiveness.

\paragraph{Base Scenario - no EV control}

Fig. \ref{fig:power_increase_no_battery} shows how an increase in production equally leads to higher values of exceedance and potential switching events. However, to what extent this happens, strongly depends on the chosen type of network. A concise classification of networks with respect to their robustness towards fluctuations will be an interesting research problem for future work. 

In this base scenario the EVs do not participate in frequency control. However, by measuring how many times $f_{min}$ was crossed, the potential switching events are determined. In order to challenge our EV grid control, as previously mentioned, we have picked a case of relatively high production, $S_i=0.268$MW (marked by black diamonds for each network in Fig. \ref{fig:power_increase_no_battery}). Fig. \ref{fig:time_series_basescenario} shows the frequency evolution for all $100$ network nodes for this model setup. The frequency safety band illustrates how much time the nodal frequencies spend outside the given safety band of $\pm20$mHz. The grey dotted lines show where the EV control would be triggered, if enabled. The dead band of $\pm10$mHz is in accordance with the present frequency regulation scheme where primary control kicks in \cite{transmission_code}.

\paragraph{How to avoid demand synchronization catastrophes with EV ramping.}

The advantage of EVs for grid control, compared with devices that have a fixed runtime, is the possibility to smoothly ramp control up and down at any time. The need for battery charging is left to an investigation of secondary and tertiary grid control. In this work, the focus is on primary control balancing of short-term fluctuations centered around a zero frequency deviation mean value.

In the following, different ramping parameterizations are tested. As the control performance is probably very sensitive towards the chosen ramping slope, in the following we vary $f_{max}$ and keep $f_{min}=0.01$Hz fixed (see Fig. \ref{fig:ramping_scheme}). 
Fig. \ref{fig:ramping_slope_increase} shows how, for an ensemble of $15$ networks, different slopes perform with respect to the number of switching events, that happen on average at each node and how many times a node on average exceeds the given frequency threshold band. In the steady state, $\dot{\omega}_i=0$, the latter mean exceedance can be related to the global frequency deviation $\Delta F$ which again can be defined as a function of $f_{max}$.  By summing over all indices $i$, we get

\begin{align*}
0 &= \sum_{i=0}^{N-1}\frac{1}{H_i} 
 \left[\frac{}{} \delta P_i- 2\pi \alpha_i\Delta F \right. \\
&- \left. (\Delta F -f_{min})\frac{p_{max}}{f_{max}-f_{min}}\right]
\label{eq:DeltaF}
\end{align*}

%\begin{eqnarray}
%\Delta F \left(\frac{N p_{max}}{f_{max}-f_{min}}+\sum_i 2\pi\alpha_i/H_i\right) \\
%= \sum_{i}\frac{\delta P_i}{H_i}+f_{min}\frac{N p_{max}}{f_{max}-f_{min}}
%\label{eq:DeltaF}
%\end{eqnarray}
%
%\begin{eqnarray}
%\Delta F = \frac{\Delta P +f_{min}\frac{N p_{max}}{f_{max}-f_{min}}}{\left(\frac{N p_{max}}{f_{max}-f_{min}}+\sum_i 2\pi\alpha_i/H_i\right)}\\
%\label{eq:DeltaF}
%\end{eqnarray}
Thus, the shape of the exceedance over $f_{max}$ function can be easily reproduced analytically (shown in blue in Fig. \ref{fig:ramping_slope_increase} (center)). 
\begin{equation}
\Delta F = \frac{\Delta P/H(f_{max}-f_{min})+c \cdot f_{min}}{d(f_{max}-f_{min})+c}
\label{eq:DeltaF}
\end{equation}
where $c = N p_{max}$, $d=\sum_i 2\pi \alpha_i/H_i$ and $\Delta P=\sum_{i=1}^{N-1}\delta P_i$ is the absolute power mismatch in the grid with contributions $\delta P_i$ from all nodes but the slack bus, $i=0$, and thus $H=H_i, \forall i={1,..,N}$. With the probability distribution of $\Delta P$ values over time a $\Delta F$-distribution could be derived and the integral over all values above $\Delta F=0.02$Hz would result in the exceedance values. 
From Fig. \ref{fig:ramping_slope_increase} (center) we conclude that at least $f_{max}=0.02$Hz is necessary to reduce the exceedance probability to zero. Compared to other work, our control scheme does not lead to an increased probability of large frequency peaks \cite{tchuisseueffects}.

The switching events are relatively insensitive to a variation in ramping slope. Only for $f_{max}=0.01$Hz where batteries are charging and decharging with an infinitely large ramping slope, switchings shoot up to more than $500$ per minute. 
Nevertheless, in the zero exceedance range of $0.01<f_{max}<0.02$Hz the number of switching events (with a mean of about $18/min$) is very high. 
Fig. \ref{fig:ramping_slope_increase} (right) illustrates why this is the case. The frequency is fluctuating around the battery treshold because all batteries react almost simultaneously to the treshold crossings. Small differences in local frequency signals are not enough to prevent the build-up of such an undesirable feedback.\par

\paragraph{How to ensure sustainable battery operation with EV randomization.}

\begin{figure}[b!]
\includegraphics[height=0.17\textwidth]{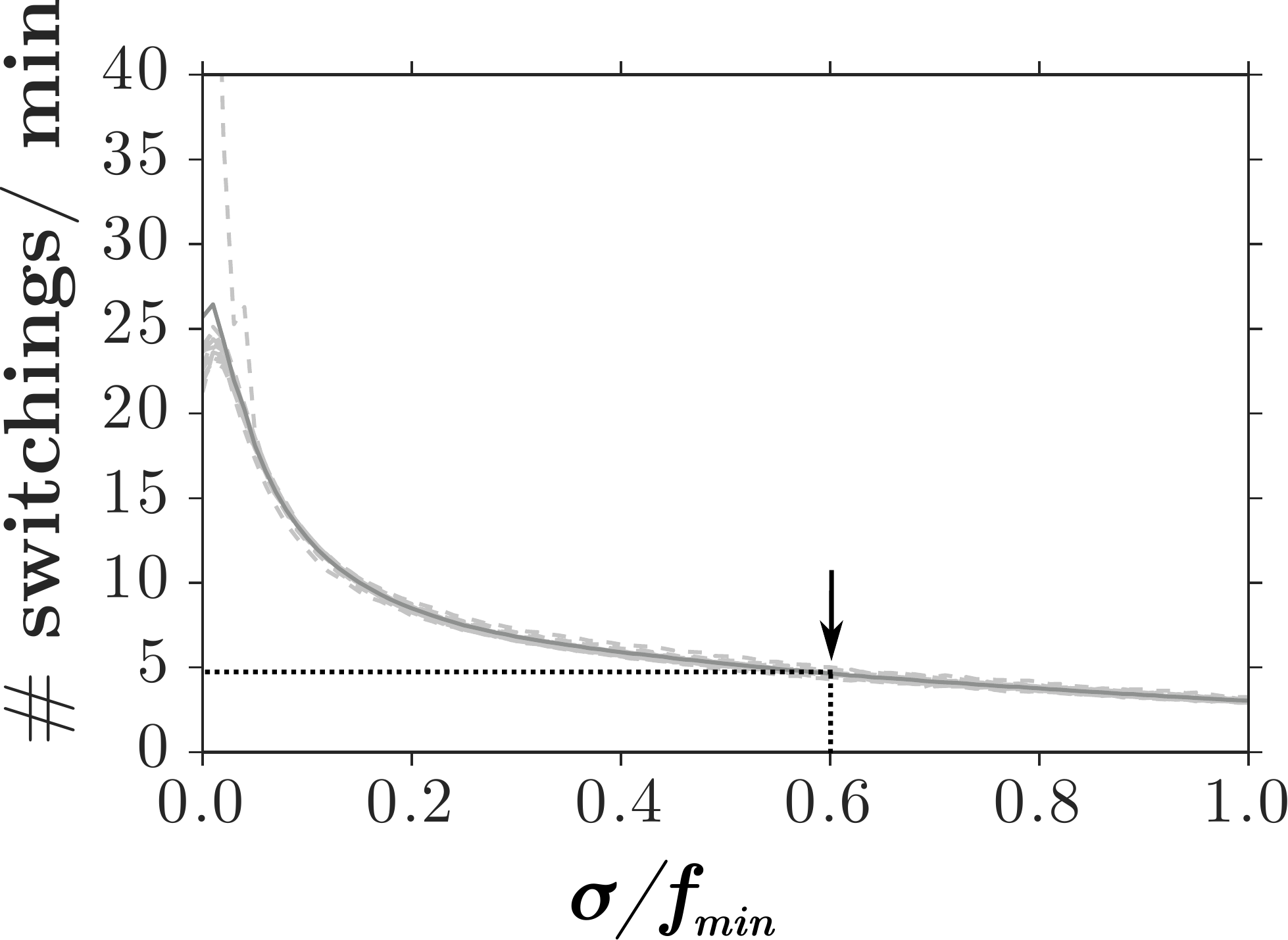}%\includegraphics[width=0.26\textwidth]{input_power0_disp52_net12_edited.pdf}
\hspace{0.01\textwidth}
\includegraphics[height=0.165\textwidth]{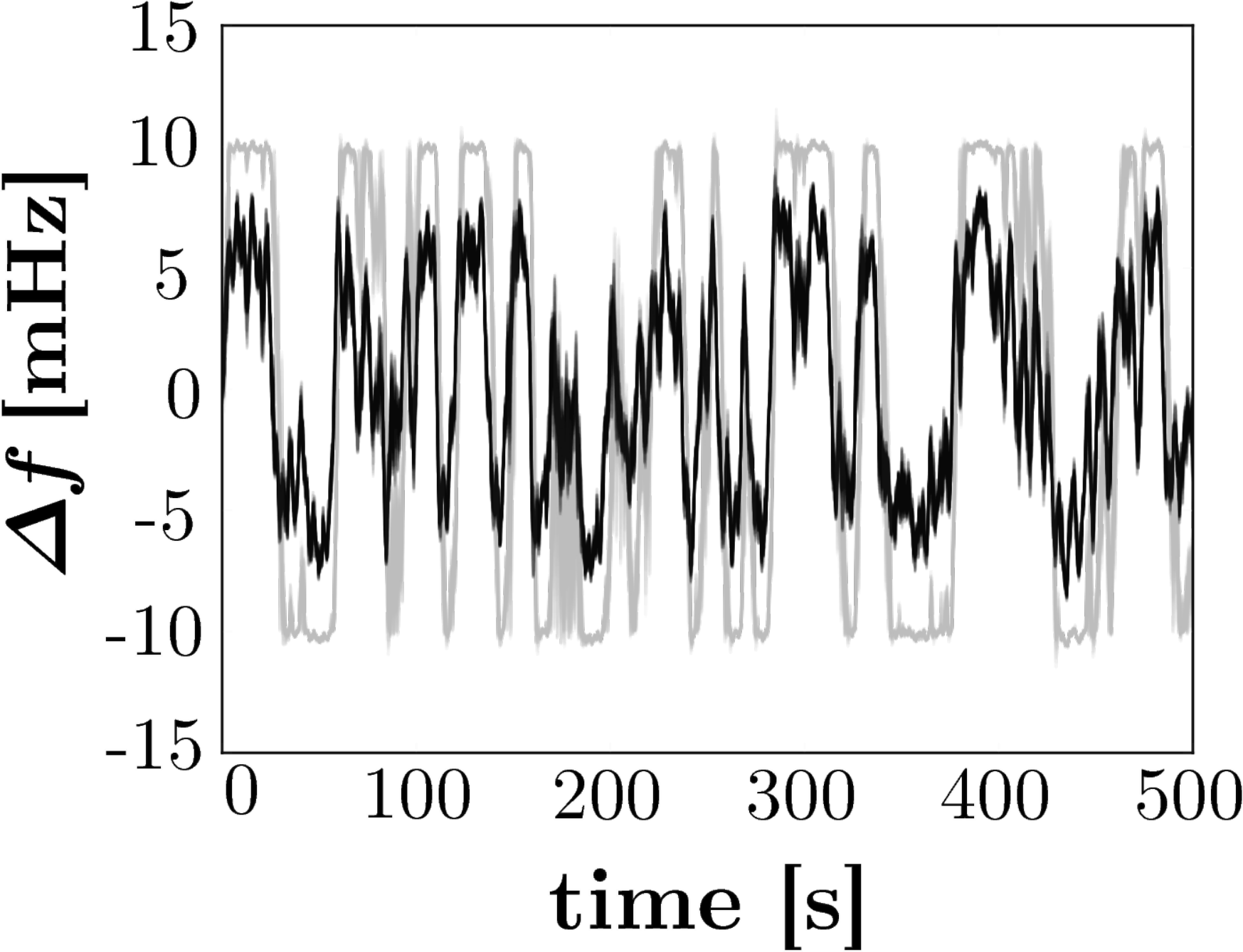}
%\hspace{0.02\textwidth}
\caption{\textbf{Reduced switchings through randomization.} Left: Number of switching over normalized variance of the battery threshold $f_{min}=10$mHz for a $15$ network ensemble. Right: Frequency trajectories for all nodes of an example network for a normalized variance of 0\% (grey) and 60\% (black).}
\label{fig:randomization}
\end{figure}

To reduce switching events, as previous work suggested, we want to randomize battery thresholds $f_{min}$ \cite{short2007stabilization,mohsenian2010autonomous,moghadam2016distributed}.
From Fig. \ref{fig:ramping_slope_increase} we have seen how a high ramping slope is able to push exceedance down to zero. However, an undesirably large number of switching events exists which would lead to fast battery degradation.
Thus, we draw the battery threshold for each EV from a Gaussian probability distribution centered around $\bar{f}_{min}=0.01$Hz. With this randomization, we prevent all EVs from switching on and off at the same time which leads to a negative feedback and oscillations around the battery threshold.
Fig.\ref{fig:randomization} (left) demonstrates how the switching events at first peak for very small variance and then rapidly decrease. Already for 20\% of variance, switching events are reduced by around 60\%, for 60\% variance they are down to 20\% of its value without randomization. The power input evolution over time reveals another side effect. In addition to the switching events also the peaks in absolute power changes are reduced. Finally, in frequency trajectories with randomization there are no oscillations around the battery threshold anymore and a normal distribution of the battery threshold around $0.01$Hz results in frequency fluctuations much below this value (see Fig.\ref{fig:randomization} right). 

Because we still want to keep a dead band for all EV batteries, in the following we will choose the 60\% variance as the standard model setup. For larger variances an increasing number of EVs would already start their control at very small frequency deviations or even close to zero. 

\begin{figure}
\includegraphics[height=0.155\textwidth]{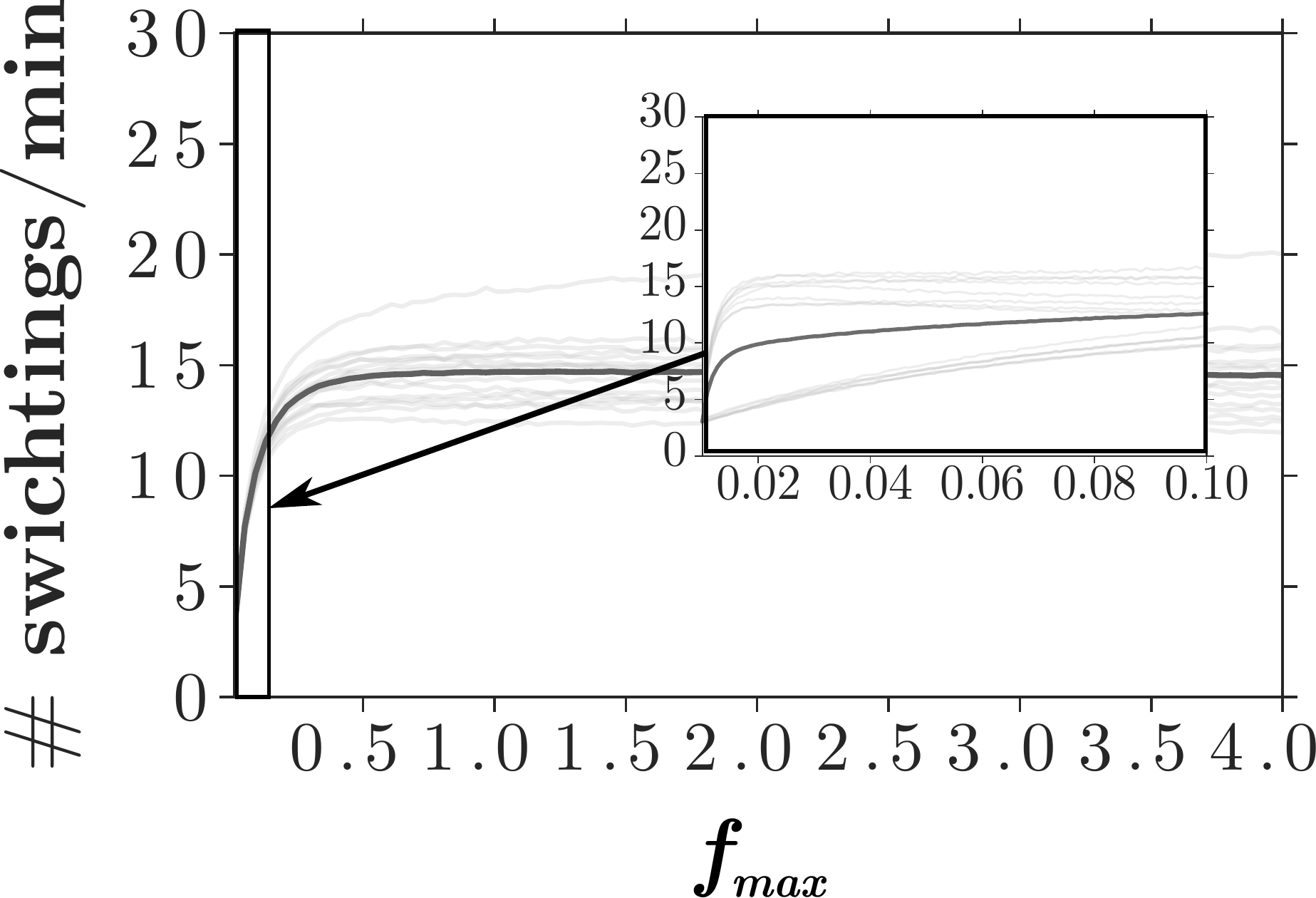}
\hspace{0.01\textwidth}
\includegraphics[height=0.155\textwidth]{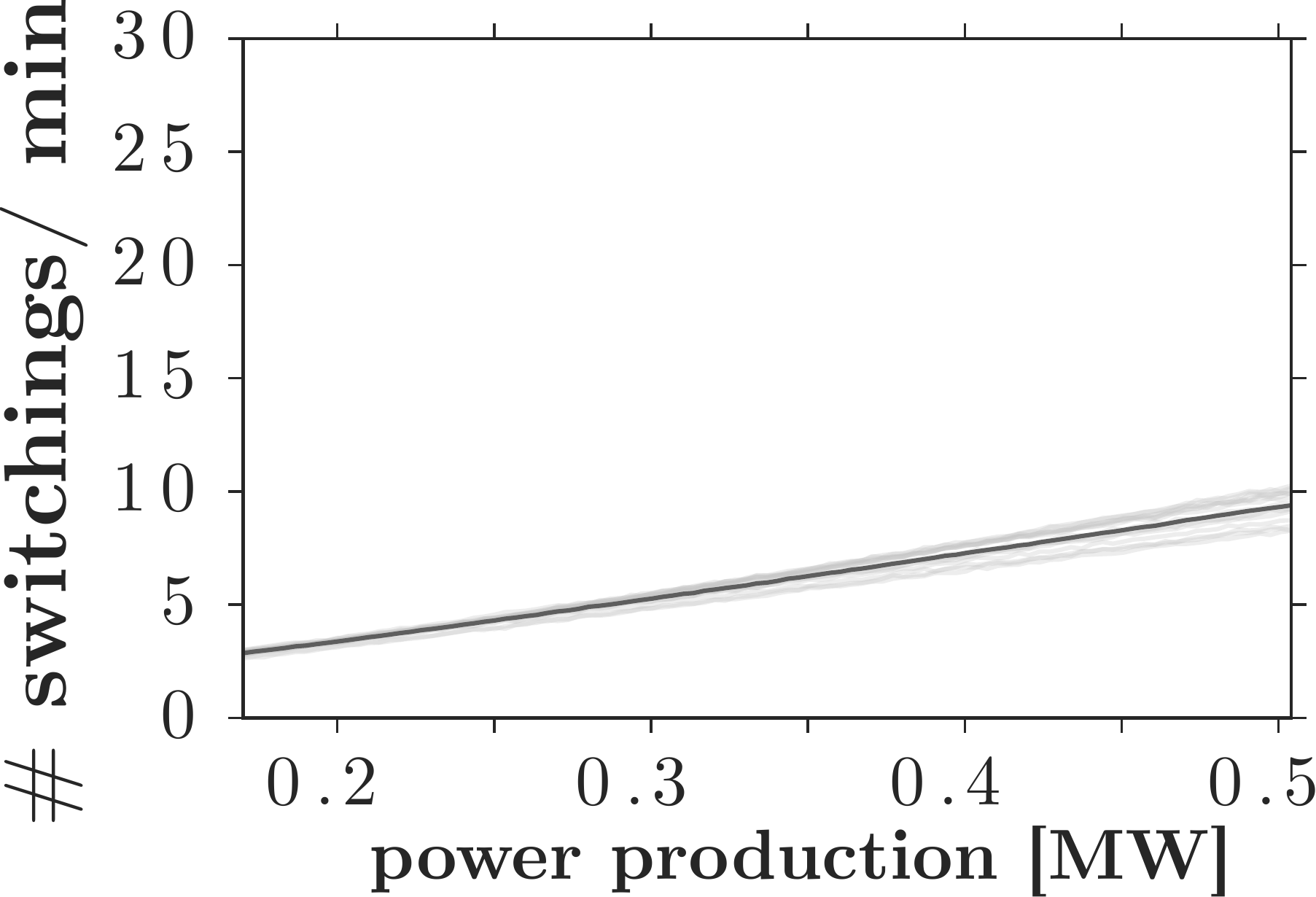}
\caption{\textbf{Robustness of the chosen EV control scheme}: Number of switchings for varying ramping slopes (left) and for increasing power production (right) with 60\% normalized variance in $f_{min}$.}
\label{fig:robustness_rand}
\end{figure}

\begin{figure}[b!]
\includegraphics[width=0.23\textwidth]{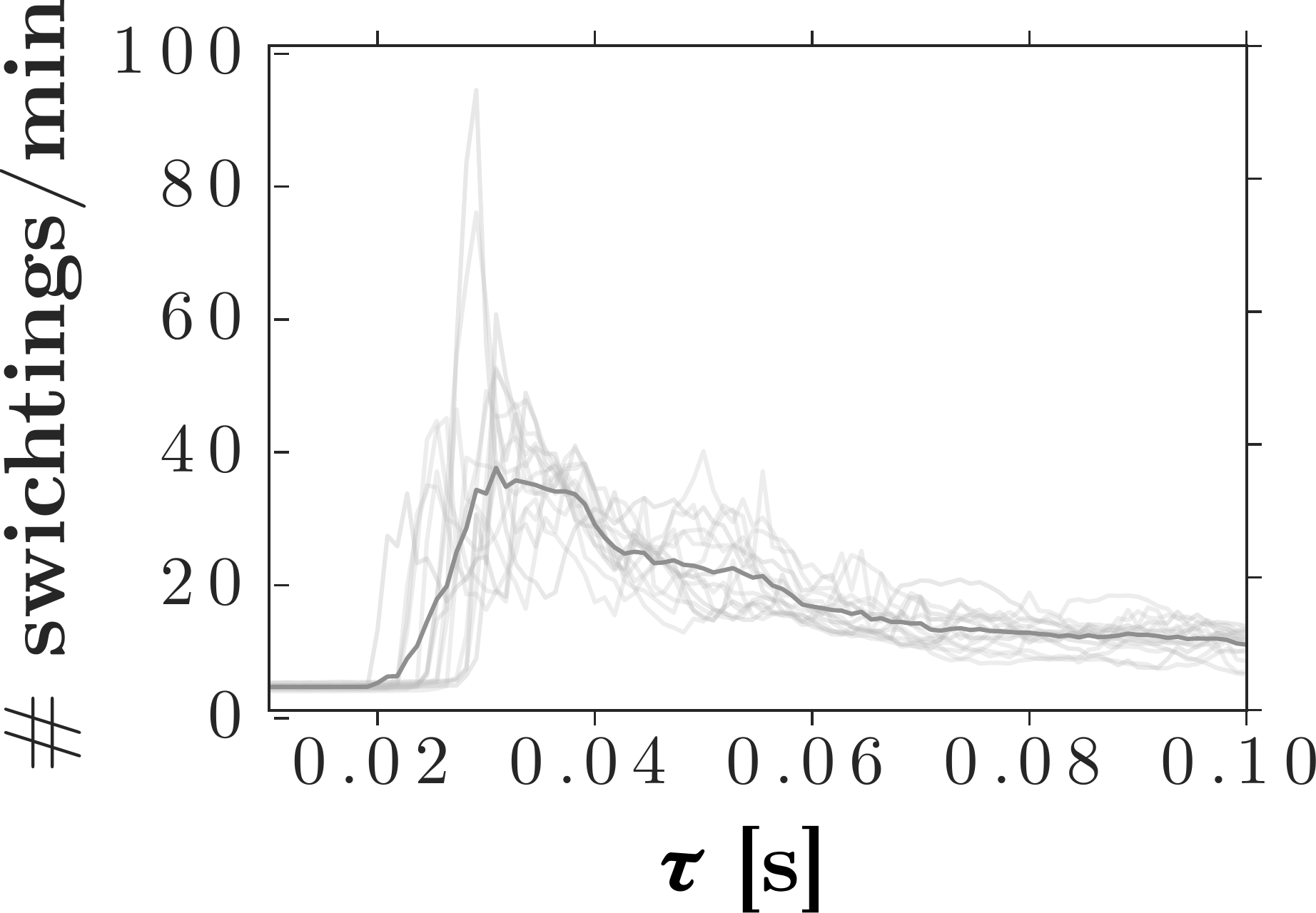}
\includegraphics[width=0.23\textwidth]{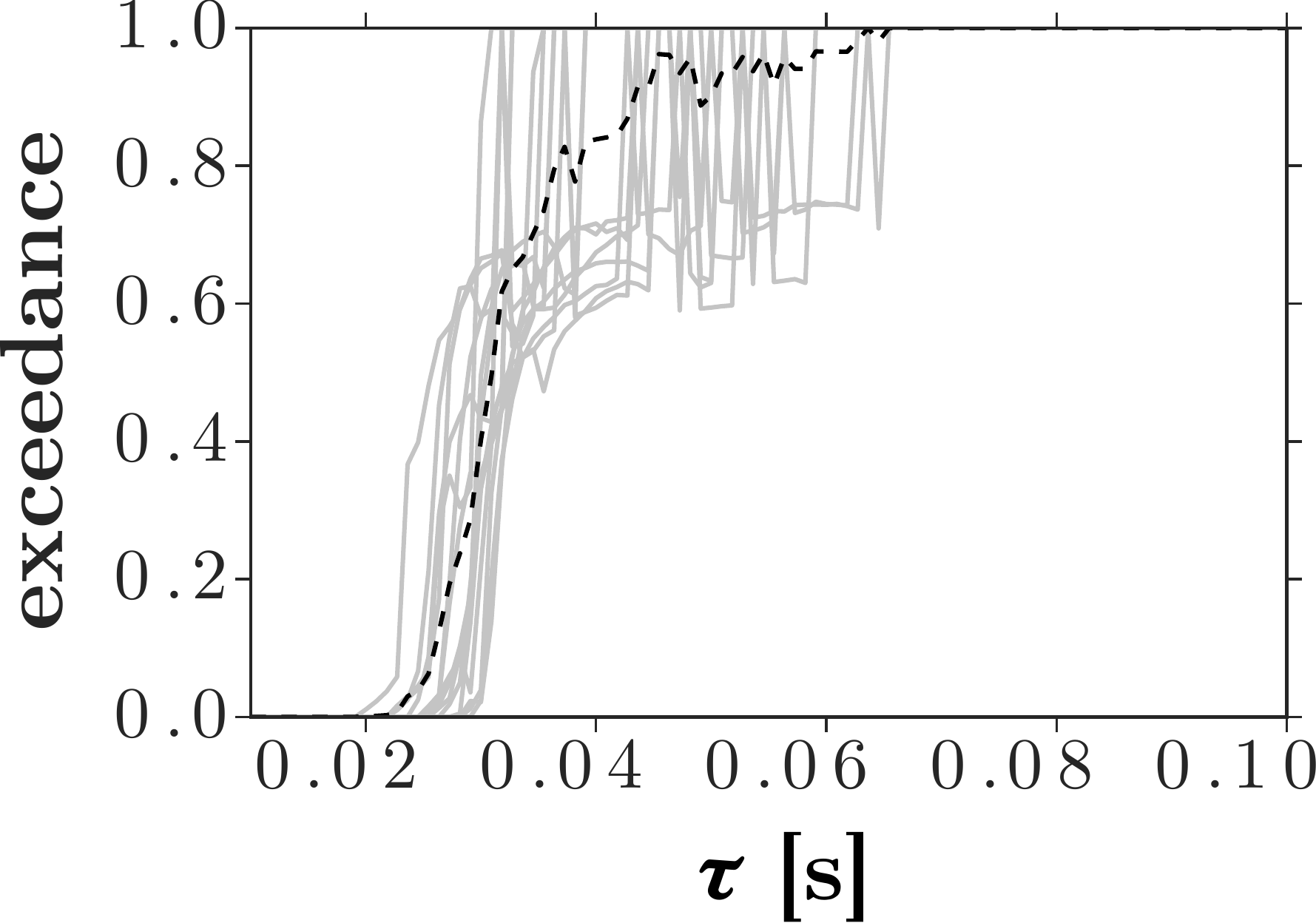}
\caption{\textbf{Influence of input signal averaging}: Averaging over the input frequency signal for different interval lengths $\tau$ (or averaging times) changes exceedance (right) and switching (left).}
\label{fig:averaging}
\end{figure}

\begin{figure*}[t!]
\includegraphics[width=0.24\textwidth]
{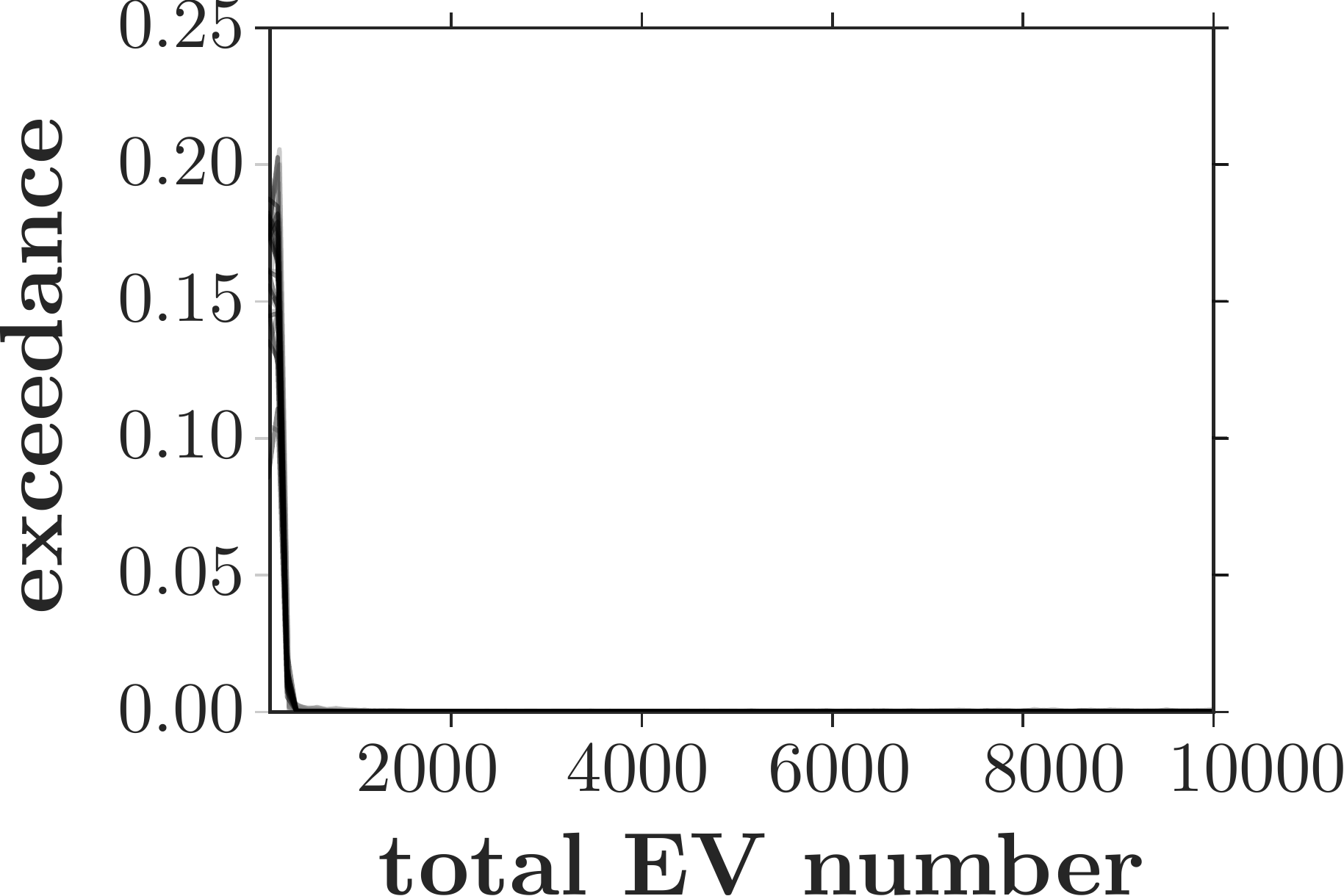}
\includegraphics[width=0.24\textwidth]
{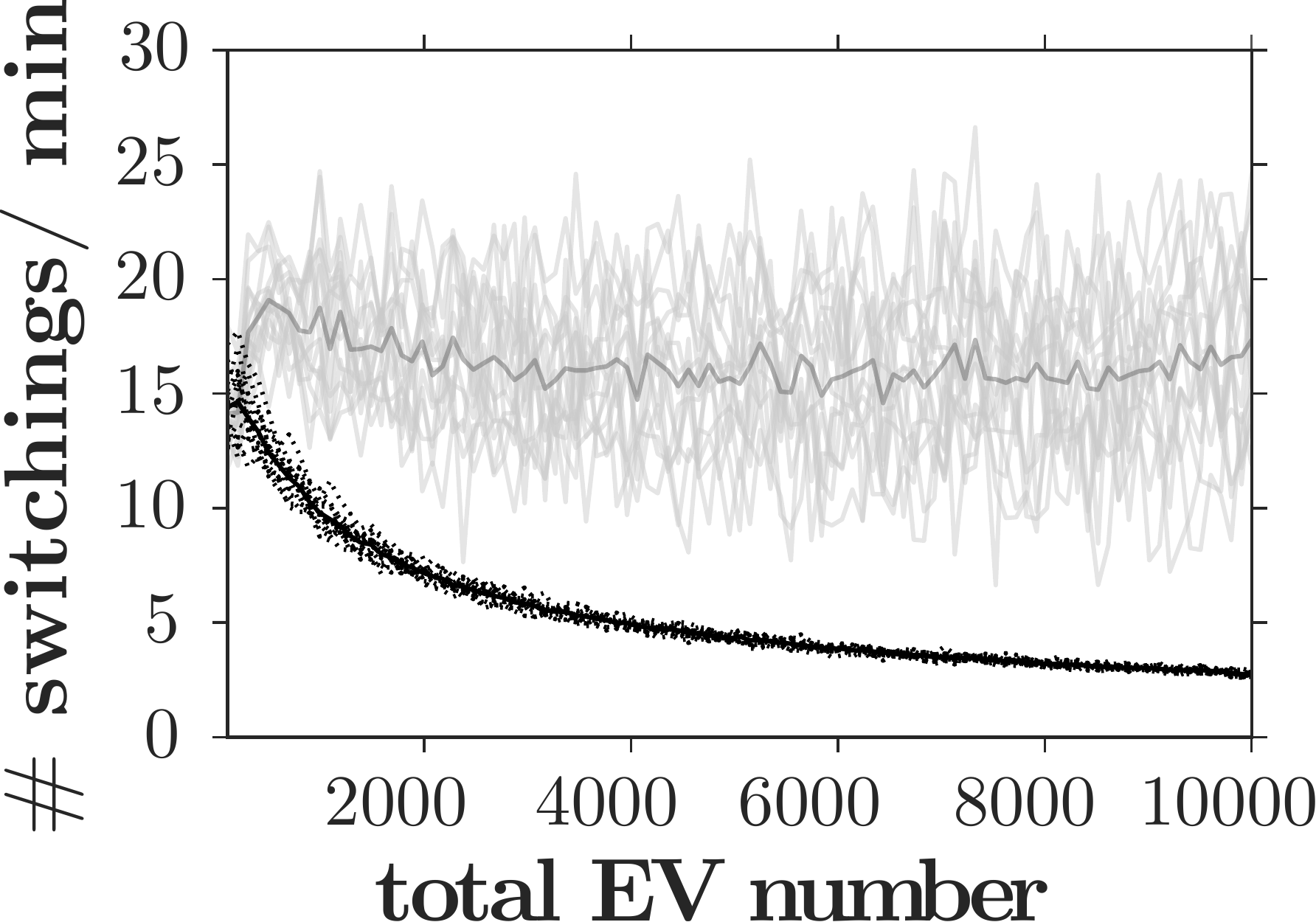}
\includegraphics[width=0.22\textwidth]{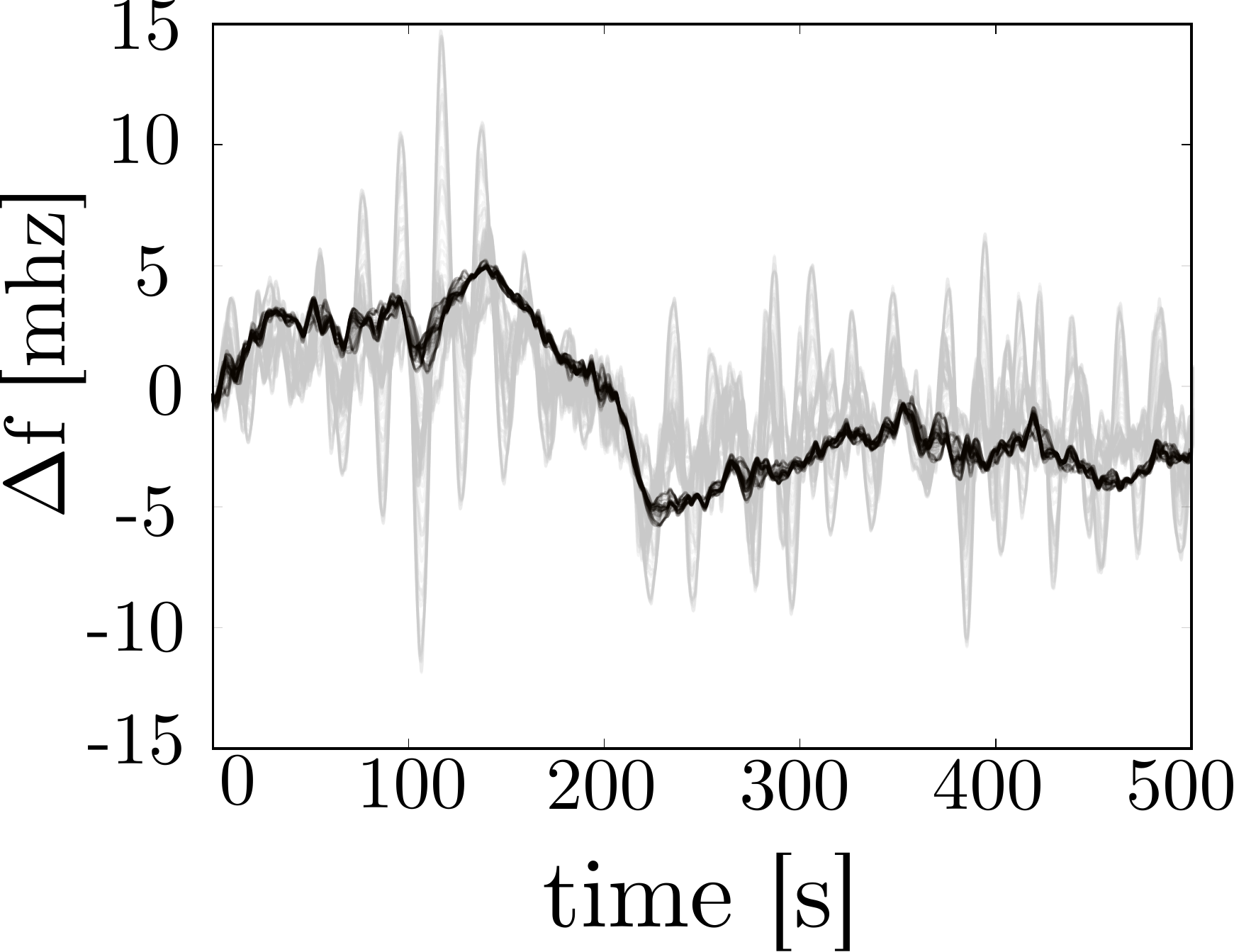}
\includegraphics[width=0.23\textwidth]{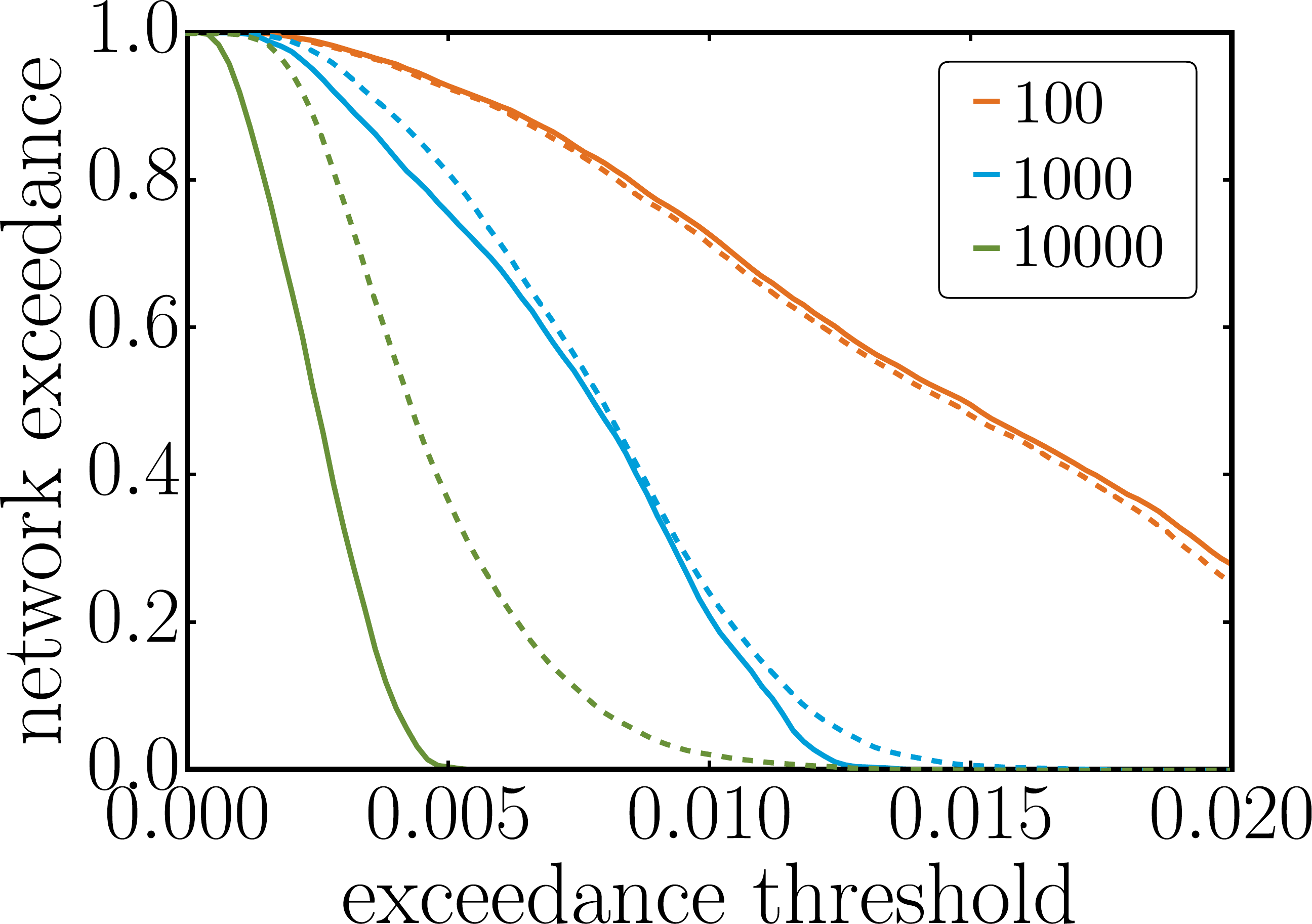}
\caption{\textbf{Decentral vs central control}: Single node exceedances (left) and nodal average of battery switching events (mid-left) for an increase in the total number of EVs for a homogeneous (black dotted lines) and an inhomogeneous (grey solid) EV distribution. The mean values are plotted with a darker color gradient. In the decentral case all nodes have the same number of EVs, whereas in the centralized approach each node but the heavy node has only 1 EV, all the other EVs are connected to the heavy node. Mid-Right: Exemplary $50$s time frame of frequency trajectories (from $100$-node example network) of an overall simulation time of $500$s for the decentral (black) and central (grey) EV distribution for the same total number of $10,000$ EVs. Right: Varying exceedance threshold and overall network exceedance for central (dashed) and decentral (solid) control and different total EV number (see legend) for an example grid.}
\label{fig:powerincrease}
\end{figure*}

\begin{figure}[b!]
\centering
\includegraphics[width=0.4\textwidth]{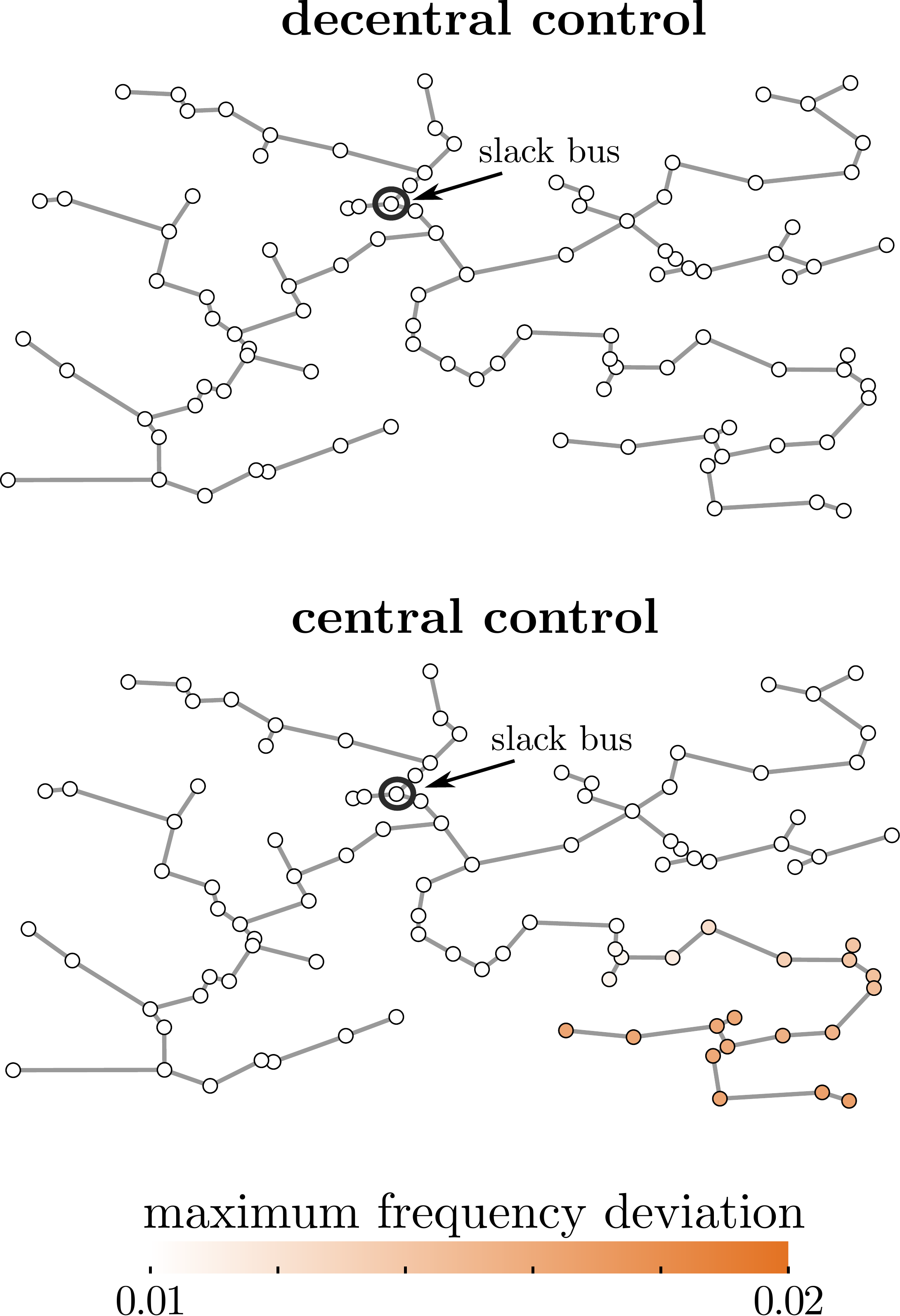}
\caption{Maximum nodal frequency deviation for decentral (top) and central (bottom) EV control for an example grid with a total of $10,000$ EVs. In the decentral case all nodes have the same number of EVs, whereas in the centralized approach each node but the heavy node has only 1 EV, all the other EVs are connected to the slack bus.}
\label{fig:network_max_freq}
\end{figure}

\paragraph{Robustness of control setup}

With this choice of EV control parameterization, the ramping slope and input power variation is repeated in order to check for any improvements. Indeed, for immediate ramping (with infinite ramping slope) the strong repeated switching is suppressed and reduced by a factor $100$ (see Fig. \ref{fig:robustness_rand} left). However, in the evolution of exceedance over $f_{max}$ (not shown here) there are no changes. According to Fig. \ref{fig:robustness_rand} (right) even with the randomization approach a light increase in switching events is  unpreventable. At the same time, with respect to the frequency exceedance, the model setup is pretty robust towards increasing power fluctuations.

\paragraph{Destabilizing effect of input signal averaging.} As an alternative to randomization, an input signal averaging approach was considered in order to reduce battery switchings.  The influence of averaging on exceedance and switchings, without any randomization present is shown in Fig. \ref{fig:averaging} (left and center). 
At averaging times around $\tau=0.02$ s the frequency fluctuations grow in time. Not only that the frequency safe-band is exceeded but the frequency is completely driven out of its stable state. Normally distributing the battery threshold around $f_{min}=0.01$ Hz does not eliminate the destabilizing effect of averaging.

We suspect that this is due to the introduction of delays into the system \cite{schafer2015decentral,schafer2016taming,yu2016use}, the further study of which is outside the scope of this work.

\paragraph{How to ensure effective control -- Central vs. decentral EC control}

Now we set up a control system that both brings down the exceedance to zero and reduces switching events through randomization. In the following, we want to test the robustness of our proposed control scheme against a changing number of EVs in the power system and compare how central vs. decentral control performs. This is realized by either distributing a number of $M$ EVs homogeneously or inhomogeneously in the power grid. In the decentral case all nodes have the same number of EVs whereas in the centralized approach all nodes except the slack bus have only one EV. The number of EVs place at the slack bus is then $M-(N-1)$.
It can be positively emphasized that both regional distributions are able to bring down exceedance when there are more than a total of about $400$ EVs in the system (see Fig. \ref{fig:powerincrease} left). Thus, above a minimum number of EVs the exceedance of the $0.02$ Hz-threshold is independent of the way EVs are distributed. However, concerning the switching events, we see a considerable difference in performance of both model cases. For the central distribution the mean switching number is one order of magnitude higher than for the homogeneous distribution and the variance in the performance for different networks is very large, as Fig. \ref{fig:powerincrease} (mid-left) shows. 

The frequency trajectories for either a central or decentral distribution of a total of $10,000$ EVs for an exemplary time frame of $50$s is illustrated in Fig. \ref{fig:powerincrease} (mid-right). The frequency fluctuations for the central case are up to three times larger for a few nodes. Because all $100$ nodes' frequency trajectories are plotted, these large fluctuations may be attributed to certain nodes in the network. 
 Indeed, it is clearly visible how the decentral EV control is able to better equally reduce frequency fluctuations among all network nodes, whereas the central control scheme is not able to handle frequency fluctuations at nodes further away from the slack bus. Fig. \ref{fig:network_max_freq} compares both cases by illustrating the maximum frequency deviation for each node over the whole time horizon in different coloring.  For stricter exceedance thresholds this would lead to notable differences in exceedance values for high numbers of EVs (see Fig. \ref{fig:powerincrease} (right)).

%\paragraph{Switching event reducing strategy}
%
%\begin{figure}[h!]
%\includegraphics[width=0.5\textwidth]{lineplot_switching_events_disp37_random_net_12.pdf}
%\includegraphics[width=0.5\textwidth]{lineplot_exceedance_disp37_random_net_12.pdf}
%\caption{Line plot of switching events per battery and exceedance for different EV number for a switching event reducing strategy.}
%\end{figure}

\section*{Conclusion}

In this work, we demonstrated the feasibility and advantages of decentral EV control for MV network ensembles with high shares of solar production and thus, strong intermittent power fluctuations. Here, we incorporated a highly realistic stochastic representation of RES fluctuations and focused on the issue of primary control of short-term frequency fluctuations centered around a mean of $50$Hz, the stable set point of frequency synchronization. We explicitly model the network structure instead of following a copper-plate approach which allows us to compare the performance of decentral vs central control by modeling the interaction of all EV devices via the power grid infrastructure.  

In our analysis, we followed the three main aims of ensuring dynamic grid stability within a frequency safe band, engineering EV control for a sustainable battery operation and designing grid control in an effective manner.
In order to ensure grid stability, we find a maximal necessary (critical) ramping rate to completely suppress threshold exceedance. The influence of the ramping rate on the exceedance can be reproduced analytically. The ability of battery devices to be adjustable in their ramping as well as charging and decharging times prevents an undesired synchronization catastrophe caused by negative feedback loops. Hence, our suggested control scheme does not lead to an increased probability of large frequency peaks \cite{tchuisseueffects}.

Nevertheless, using the same ramp for all EVs leads, as expected, not to a synchronization catastrophe but still to a synchronization of the control devices with the result of
 a large number of battery switching events. To overcome this effect and prevent battery degradation, we introduce a variance in battery threshold and randomize the switching of the different EV devices. 
To our knowledge, this combination of ramping slope and randomized battery threshold, performs best to jointly reduce exceedance and switching events. This control strategy parameterization is relatively robust against a further increase in power production and thus fluctuations. The exceedance stays at zero level and switching events only increase slightly.

In contrast to the randomization, an averaging approach destabilizes the system. This highlights the need for further research into the interaction of decentral frequency control and delayed control actions.

Another important finding of this paper is the advantage of decentral over central control for a more effective frequency balancing. While both control measures succeed in keeping fluctuations within a given safe band, the decentral control leads to an order of magnitude lower switchings and thus, allows for a more sustainable battery operation. At the same time, the central control would introduce a strong heterogeneity in  fluctuation amplitudes among the network nodes.

For further work, we see a great potential in the extension of our model setup to secondary and tertiary control. This would also allow to incorporate EV control into a realistic case study and compare it with other balancing techniques with respect to their technical and economic feasibility.  Related to this issue is the interaction of EV control with different inverter types and their individual control schemes.

Generally, electric vehicles are an opportunity for the decarbonization of both the electricity and traffic sector, especially by interconnecting the two. The use of state-of-the-art battery technology increases the availability of storage for the eradication of mid- and short-term power fluctuations, e.g. from RES deployment. With our holistic network modeling approach, we demonstrated the technical feasibility of interconnected EV control devices but, there is much more work to follow to understand the risks and potential of decentral EV grid control.

\section*{Acknowledgement}
S.A. wants to thank her fellow colleagues Paul Schultz and Anton Plietzsch for helpful discussion and comments.
The authors gratefully acknowledge the support of BMBF, CoNDyNet, FK. 03SF0472A and the European Regional Development Fund (ERDF), the German Federal Ministry of Education and
Research and the Land Brandenburg for supporting this project by providing resources on the high performance computer system at the Potsdam Institute for Climate Impact Research.

% Either type in your references using
% \begin{thebibliography}{}
% \bibitem{}
% Text
% \end{thebibliography}
%
% or
%
% Compile your BiBTeX database using our plos2015.bst
% style file and paste the contents of your .bbl file
% here.
% 

\bibliographystyle{plos2015}
\bibliography{bibfile.bib}%.bib

% that's all folks
\end{document}